# 0.6-V, µW-Power 4-Stage OTA with Minimal Components and 100X Load Range

M. Privitera, *Member, IEEE*, A. D. Grasso, *Senior Member, IEEE*,
A. Ballo, *Member, IEEE*, and M. Alioto, *Fellow, IEEE*

*Abstract*— A four-stage operational transconductance amplifier (OTA) for ultra-low-power applications is introduced in this paper. The proposed circuit inclusive of frequency compensation requires minimal transistor count and passives, overcoming the traditionally difficult compensation of 4-stage OTAs and bringing it back to the simplicity of 3-stage OTAs. At the same time, the proposed circuit achieves high power efficiency, as evidenced by the >3.7X (>11.3X) improvement in the large-signal (small-signal) power efficiency figure of merit $FOM_L$ ($FOM_S$), compared to prior 4-stage OTAs (sub-1 V multi-stage OTAs). Thanks to the lower sensitivity of the phase margin to the load capacitance, the proposed OTA remains stable under a wide range of loads (double-sided as in any 3-4-stage OTA), achieving a max/min ratio of the load capacitance of >100X.

## I. INTRODUCTION

THE prioritization of digital over analog in the optimization of commercial CMOS processes has inevitably made analog design more challenging [1], [2]. The reduction in the transistor intrinsic gain in modern CMOS technologies[1] in low-power systems mandates a larger number of gain stages in building blocks requiring high gain, such as operational transconductance amplifiers (OTAs). 3- and 4-stage amplifiers provide high DC gain and closed-loop gain accuracy at reduced bandwidth, due to the presence of additional low-frequency poles. For the same reason, closed-loop stability requires the load to be simultaneously lower- and upper-bounded with limited load range [3]-[11] (i.e., maximum over minimum load capacitance $C_{L,max}/C_{L,min}$). This restricts design reuse of the same OTA design IP across on-chip building blocks with different loads. At the same time, this limitation becomes particularly problematic in applications where the same OTA load experiences large fluctuations during its operation [6], [7], [9]-[11]. Examples include switched-capacitor circuits, line drivers, MEMS, headphone drivers, among many others [11].

In $N$-stage OTAs, frequency compensation can be achieved via classical nested Miller compensation (NMC) with at least $N − 1$ compensation capacitors. While intrinsically simple, each capacitor in NMC is proportional to the load capacitance $C_L$, which leads to severe bandwidth limitations and large silicon area [12]. To overcome such issues, designs based on single-Miller capacitor (SMC) have been recently proposed for three- and four-stage OTAs [9], [12]. However, their load range is currently limited to a few units [5], or very few tens [6], [9].

This paper introduces an OTA including a novel frequency compensation network with single-Miller capacitor having minimal component count across existing 4-stage OTAs. The OTA is shown to have best-in-class load driving efficiency with the lowest supply voltage and the lowest power consumption, while reaching the 100X mark for the load capacitance range.

The manuscript is structured as follows. The proposed OTA is presented in Section II, and analyzed in Section III. Section IV presents measurements. Conclusions are in Section V.

## II. PROPOSED OTA ARCHITECTURE

The simplified schematic of the proposed amplifier inclusive of frequency compensation network is illustrated in Fig. 1. The input stage is a partial folded-cascode amplifier comprising transistors $M_1$–$M_8$ biased by transistor $M_{5B}$. As a distinctive feature of the proposed circuit configuration, all the remaining stages are implemented by simple common source stages $M_9$–$M_{10}$, $M_{11}$–$M_{12}$, and $M_{13}$–$M_{14}$, respectively. This entails the minimum possible number of transistors for a four-stage OTA, and is equal to that of a conventional three-stage OTA [12]. In contrast, prior OTAs with four stages use fully-folded cascode differential pair [9], more complex non-inverting stages [6], and/or auxiliary stages in parallel to the last one [6], [9].

In Fig. 1, the gate of the active load transistor in the last stage $M_{14}$ is directly connected to the output of the first stage, forming a feed-forward class-AB output stage with improved driving capability and efficiency under heavy capacitive loads [5], [6]. In principle, two further transistors could be saved by using a simple differential pair with current mirror active load for the first stage. However, this would require higher voltage headroom and supply voltage.

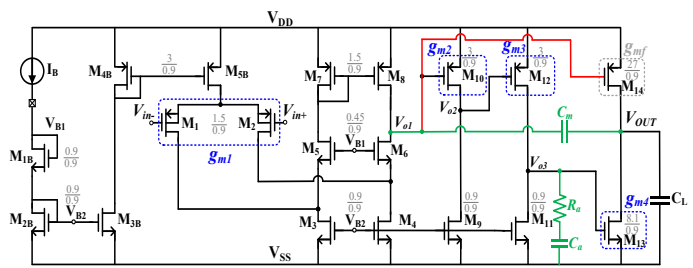

Fig. 1. Proposed OTA including passives for frequency compensation.

Manuscript received on May 19, 2024. This project has received funding from the European Union's Horizon Europe research and innovation program under the Marie Skłodowska-Curie grant agreement no. 101086359, and the MOE-T2EP50223-0016 grant from the Singapore Ministry of Education.

Marco Privitera, Alfio Dario Grasso, and Andrea Ballo and are with the DIEEI department – University of Catania, Italy (emails: marco.privitera@phd.unict.it, alfiodario.grasso@unict.it, andrea.ballo@unict.it). Massimo Alioto is with the ECE Department, National University of Singapore, Singapore (e-mail: malioto@ieee.org).

---

[1] For example, the gain per stage in 180 nm at ≤0.6 V is ~30-dB (see below, [4], [8]), and certainly lower in more recent technologies (e.g., [10]).

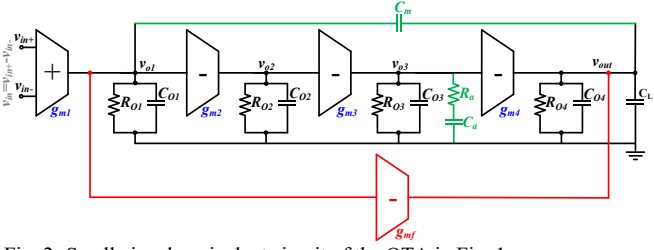

Fig. 2. Small-signal equivalent circuit of the OTA in Fig. 1.

The proposed frequency compensation in Fig. 1 is based on 1) a single Miller capacitor $C_m$ connected between the output of the first stage and the OTA output node, and 2) an $RC$ circuit ($R_a$ and $C_a$ in Fig. 1) connected to the output of the third stage. This compensation network has the minimum number of passives among 4-stage OTAs [5], [6], [9], [11]. The compensation network arrangement and proper design strategy lead to a load range of 100X as shown in Sections III-IV.

The small-signal equivalent circuit of Fig. 1 is illustrated in Fig. 2 to support circuit analysis in the following. In this figure, parameters $g_{mi}$, $R_{oi}$ and $C_{oi}$ are respectively the equivalent transconductance, output resistance and capacitance of the $i$-th OTA gain stage ($i$=1…4). The feed-forward transconductance gain $g_{mf}$ in Figs. 1-2 is associated with transistor M14.

## III. OTA CIRCUIT ANALYSIS

### A. Small-Signal Stability Analysis

In the open-loop transfer function of the circuit in Fig. 2, it is assumed that $g_{mi}R_{oi} \gg 1$ (i.e., the gain stages are reasonably large), $C_m \ll C_L$ to limit the area of passives, $R_a \ll R_{oi}$, and $C_a \gg C_{oi}$ (i.e., passives dominate over transistor parasitics). The resulting transfer function can be approximated as

$$A(s) \approx \frac{A_0}{1+\frac{s}{\omega_D}} \cdot \frac{1+a_1 s}{(1+b_1 s)(1+b_2 s+b_3 s^2)(1+b_4 s)}$$
$$\approx \frac{A_0}{1+\frac{s}{\omega_D}} \cdot \frac{1}{(1+b_2 s+b_3 s^2)} \quad (1)$$

where $\omega_D$ is the dominant pole, $A_0$ is the open-loop DC gain, and the transfer function coefficients are found to be

$$\omega_D = \frac{1}{g_{m2}g_{m3}g_{m4}R_{o1}R_{o2}R_{o3}R_{o4}C_m}, \quad (2)$$

$$A_0 = g_{m1}g_{m2}g_{m3}g_{m4}R_{o1}R_{o2}R_{o3}R_{o4} \quad (3)$$

$$a_1 \approx b_1 = R_a C_a \quad (4)$$

$$b_2 = \frac{C_L}{g_{m2}g_{m3}g_{m4}R_a R_{o2}} \quad (5)$$

$$b_3 = \frac{C_L C_{o2}}{g_{m2}g_{m3}g_{m4}R_a} \quad (6)$$

$$b_4 = C_{o3} R_a \quad (7)$$

where $1/b_4$ is assumed to be much higher than the unity gain frequency, as a consequence of the above conditions that make $b_4$ much lower than $b_1$, $b_2$ and $b_3$. Observe that the adoption of a bulk-driven input stage for expanded input common-mode range would keep the above expressions and the pole/zero relative placement (i.e., ratio) identical, since the bulk-driven option would simply reduce $g_{m1}$. Hence, all following conclusions are equally valid for bulk-driven 4-stage OTAs.

From (4) the proposed compensation network carries out pole-zero cancellation, while easily keeping the pole-zero doublet $1/R_a C_a$ well above the unity-gain frequency $\omega_{GBW} \approx A_0 \omega_D$ (i.e., $g_{m1}/C_m$ from (2)-(3)), making its effect uninfluential [9].

The transfer function in (1) contains two complex and conjugate non-dominant poles whose natural frequency and damping factor are given by

$$\omega_0 = \sqrt{\frac{1}{b_3}} = \sqrt{\frac{g_{m2}g_{m3}g_{m4}R_a}{C_L C_{o2}}} \quad (8)$$

$$\xi = \frac{b_2}{2\sqrt{b_3}} = \frac{1}{2R_{o2}}\sqrt{\frac{C_L}{g_{m2}g_{m3}g_{m4}R_a C_{o2}}} \quad (9)$$

The resulting phase margin $\phi$ is

$$\phi = 180° - tan^{-1}\left(\frac{\omega_{GBW}}{\omega_D}\right) - tan^{-1}\left(\frac{b_2 \omega_{GBW}}{1-b_3 \omega_{GBW}^2}\right) \approx$$
$$\approx 90° - tan^{-1}\left(\frac{b_2 \omega_{GBW}}{1-b_3 \omega_{GBW}^2}\right) \quad (10)$$

Regarding the load range, the above equations related to the proposed OTA and frequency compensation strategy can be used to evaluate the minimum and maximum $C_L$ that assure the desired closed-loop stability and transient behavior. Light-load stability is limited by the damping factor (9) of the complex and conjugate poles. Indeed, lower values of $C_L$ lead to low damping factors, and hence large peaking in the loop gain frequency response, or equivalently low gain margin [9]. The minimum drivable $C_L$ is therefore set by the value that makes (9) equal to the targeted damping factor (e.g., $\xi$=0.5 with minimum $C_L$<10 pF). Regarding the maximum $C_L$, the phase margin in (10) decreases at higher load capacitance, and it is obtained from (10) by setting a desired phase margin target (e.g., 45°). The above considerations can be reversed to further reduce (increase) the minimum (maximum) $C_L$ allowed. The feed-forward transconductance $g_{mf}$ in Figs. 1-2 has an insignificant effect on closed-loop stability, and its benefits on the large-signal response are discussed in the following subsection.

### B. Large-Signal Analysis

The slew rate ($SR$) of the OTA is constrained by the currents charging the capacitive loads at the output of each stage. Being $I_1$, $I_2$, $I_3$ and $I_4$ the maximum (dis)charge currents at the output of the four stages, the overall $SR$ is given by

$$SR = \min\left(\frac{I_1}{C_m+C_{o1}}, \frac{I_2}{C_{o2}}, \frac{I_3}{\frac{R_{o3}C_{o3}}{R_a C_a + R_{o3}C_{o3}}C_a}, \frac{I_4}{C_L}\right) \quad (11)$$

Choosing $C_a$ such that $R_a C_a \gg R_{o3} C_{o3}$ and $C_m \gg C_{o1}$, (11) simplifies into

$$SR \approx \min\left(\frac{I_1}{C_m}, \frac{I_2}{C_{o2}}, \frac{I_3}{\frac{R_{o3}}{R_a}C_a}, \frac{I_4}{C_L}\right) \quad (12)$$

In (12), the transient output current $I_4$ in the fourth stage includes the dynamic current of M14 in Fig. 1, which is much higher than its quiescent current thanks to the feed-forward gate connection to the output of the first stage, instead of a bias source. Accordingly, the large-signal behavior of the OTA with frequency compensation strategy in Fig. 1 is not restricted by the driving capability at the output for low to moderate values of $C_L$. Accordingly, the proposed OTA is expected to have a favorable tradeoff between power consumption and driving capability, as shown in Section IV. Of course, the slew rate becomes again limited by $C_L$ for heavier capacitive loads.

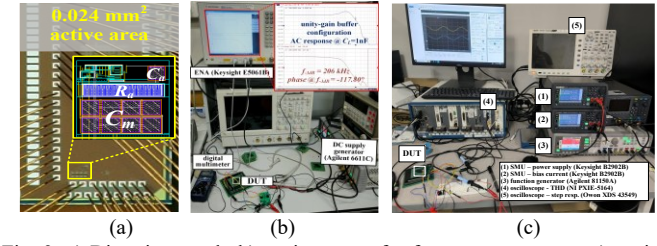

Fig. 3. a) Die micrograph, b) testing setup for frequency response, c) testing setup for time-domain characterization and accurate power measurement.

## IV. MEASUREMENT RESULTS

The OTA was designed in 180 nm with transistor aspect ratios as in Fig. 1. From Fig. 3a, its area is 0.024 mm² and is dominated by the compensation network ($C_m$=10.5 pF, $C_a$=1.2 pF, $R_a$=200 kΩ). From an area viewpoint, transistors were sized for $g_m/I_D$=23.5 V⁻¹ (hard to compare, as prior art does not report the design target), under the sub-threshold bias current $I_B$=200 nA. All transistor lengths are five times the minimum allowed by the technology for gain and bandwidth trade-off. The OTA was tested at 0.6 V.

The dynamic open-loop response of the OTA is shown in Figs. 4a-b under a load capacitance $C_L$ of 10 pF and 1 nF. The minimum (maximum) $C_L$ maintaining an acceptable phase margin is <10 pF (1 nF), allowing a wide load range of >100X. In the following, the minimum $C_L$ is pessimistically assumed to be 10 pF, which is the minimum measurable value (lower-bounded by the pad, bonding and PCB capacitance), although in simulations it can be greatly reduced to 10 fF.

The closed-loop small-signal response to a ±25 mV step in the unity-gain configuration is depicted in Figs. 4c-d. The overshoot in the rising edge is well known to be due to the different slew rate at the output of the first and the fourth stage. In applications where such overshoot is an issue (e.g., sampled data circuits), a slew rate enhancer circuit can be adopted to eliminate it [7]. The large-signal step response is shown in Fig. 5a for 10 pF load with a slew rate of 128 V/ms and in Fig. 5b 118.5 V/ms for 1 nF load as expected from (12). Fig. 6a shows that the measured input common-mode range is upper bounded at 0.45 V under 0.6-V supply, as expected from the adoption of a PMOS differential pair. Fig. 6b shows the total harmonic distortion in the unity-gain closed-loop configuration, whose worst-case value is below 3% with an input amplitude of 200 mV, and 1.5% under a 150-mV amplitude.

Ten die samples of the OTA were characterized to evaluate the impact of process variations as opposed to most prior art (only [3] and [9] tested multiple dice), as summarized in Table II in terms of mean $\mu$ and variability $\sigma/\mu$ for key parameters. The input-referred offset across dice is 3.35 mV.

Table II shows the comparison of the proposed OTA to state-of-the-art 4-stage OTAs, as well as other high gain (>75 dB) 2- and 3-stage for completeness. The proposed OTA and frequency compensation strategy exhibits the lowest number of both transistors and compensation capacitors, compared to 4-stage OTAs [5], [6], [9], as well as 3-stage [8], and on par with 2-stage OTAs [2]-[4]. The measured load range is >100X and represents an improvement of >3.3-33X over [5], [6] and [9].

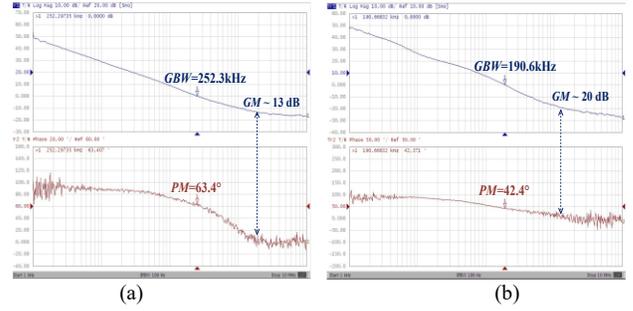

Fig. 4. Measured small-signal dynamic response of the OTA in Fig. 1: open-loop small-signal frequency response under load capacitance of a) 10 pF, b) 1 nF, and closed-loop response to ±25 mV step in unity-gain configuration under c) 10 pF, d) 1 nF.

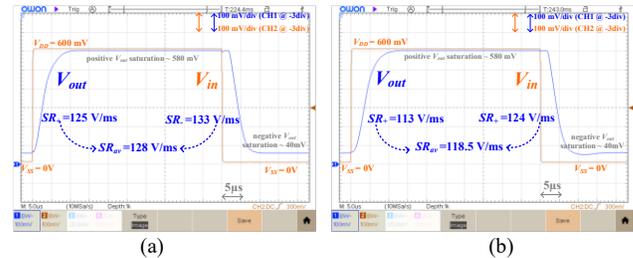

Fig. 5. Measured large-signal step response in unity gain configuration under load capacitance of (a) 10 pF and (b) 1 nF.

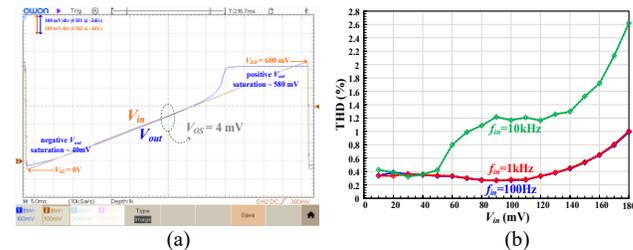

Fig. 6. Linearity in unity-gain configuration: a) input common-mode range, b) total harmonic distortion vs. peak-to-peak amplitude of input sine wave.

The measured gain-bandwidth product (Fig. 4b) $GBW$ under 1-nF load is 192 kHz when averaged across the ten dice and has a low percentage variability of 1.1%.

## V. REMARKS AND CONCLUSIONS

From Figs. 4a-b, the measured DC gain is 119.3 dB when averaged across the ten dice and has a low variability of 0.5%. In addition, the small-signal step response showed an average 1%-settling time (see Fig. 4d) of 13.5 µs, and a variability of 7.3%. The large signal performance is evaluated through the slew rate in Figs. 5a-b. The slew rate averaged across dice is 118.5 V/ms (with $C_{L,max}$=1 nF) and its variability is 1.8%. The power consumption averaged across dice is 1.65 µW with 2.8%

TABLE II. PERFORMANCE PARAMETERS AND COMPARISON WITH STATE OF THE ART (BEST PERFORMANCE OR NOTABLE PROPERTY IN BOLD)

| parameters | This work [d] | TCASI'23 [10] | TVLSI'23 [9] | TVLSI'20 [6] | TCASI'15 [5] | ACCESS'23 [8] | ACCESS'23 [11] | SSCL'22 [7] | ACCESS'20 [4] | TCASII'16 [3] | JSSC'02 [2] |
|---|---|---|---|---|---|---|---|---|---|---|---|
| technology [nm] | 180 | 65 | 65 | 130 | 350 | 130 | 65 | 65 | 180 | 180 | 500 |
| area [mm$^2$] | 0.024 | 0.013 | 0.0086 | 0.007 | 0.014 | 0.002 | **0.017** | 0.003 | 0.0098 | 0.018 | 0.5 |
| # of stages | 4 | 4 | 4 | 4 | 4 | 3 | 3 | 3 | 3 | 2 | 2 |
| # of transistors [a] | **14** | 17 | 17 | 28 | 20 | 18 | 20 | 20 | 17 | 23 | 37 |
| # of res./# of caps. | **1 / 2** | 3 / 3 | 2 / 4 | **1 / 2** | 3 / 3 | - / 3 | 2 / 3 | 2 / 3 | - / 2 | - / 2 | 2 / 2 |
| op. mode [b] | SUB, GD | SAT, GD | SAT, GD | SAT, GD | SAT, GD | SUB, BD | SAT, GD | SAT, GD | SUB, BD | SUB, GD | SUB, GD |
| $V_{DD}$ [V] | **0.6** | 1 | 1.2 | 1.2 | 3 | **0.3** | 1.2 | 1.2 | 0.3 | 0.5 | 0.9 |
| # of dice tested | 10 | 1 | 7 | 1 | 1 | 1 | 1 | 7 | 1 | 4 | 1 |
| power ($\sigma/\mu$) [µW] | 1.65 [a,d] (2.8% [e]) | 353.1 | 165.84 | 175.2 | 156 | 0.034 | 8.88 | 15.68 | 0.013 | 0.07 | 0.45 |
| gain ($\sigma/\mu$) [dB] | 119 [d] (0.5% [e]) | 90 | 132 | 107 | **173** | 86.83 | 110 | 100 | 98.1 | 77 | 79 |
| $C_{L,max}$ [pF] | 1,000 | 1·10$^8$ | 100,000 | 12,000 | 1,000 | 35 | 100,000 | 100,000 | 30 | 40 | 12 |
| $C_{L,max}/C_{L,min}$ | >100 (100,000 [c]) | **1,000,000** | 21 | 30 | 3 | N/A | 500 | 3333 | N/A | N/A | N/A |
| $GBW$ ($\sigma/\mu$) [MHz] | 0.192 [d] (1.1% [e]) | 2·10$^{-6}$ | 0.27 | 1.18 | **3** | 0.01 | 0.030 | 0.022 | 0.0031 | 0.004 | 0.006 |
| $SR_{av}$ ($\sigma/\mu$) [V/µs] | 0.118 [d] (1.8% [e]) | 1·10$^{-7}$ | 0.03 | 0.14 | **1.18** | 0.0038 | 0.010 | 0.044 | 0.0091 | 0.002 | N/A |
| $V_{OS}$ [mV] | 3.35 | N/A | N/A | N/A | N/A | 5.73 [c] | N/A | N/A | 3.1 | 4.8 | **2.6** |
| CMRR [dB] | 182 @ DC [c] | N/A | N/A | N/A | **228 @ DC** [c] | 57.80 [c] | 71.96@DC | N/A | 60 @ DC | 55 @ DC | 59 |
| PSRR ($\sigma/\mu$) [dB] | 51 [d] @ (2.2% [e]) | N/A | N/A | N/A | 110 @ DC [c] | 46.59 [c] | **182@DC** [c] | 142@DC [c] | 61 @ DC | 52 @ DC | N/A |
| noise [nV/√Hz] | 750 @ 1 kHz [c] | N/A | N/A | N/A | **724 @ 1 kHz** | 2,860 | N/A | N/A | N/A | N/A | N/A |
| $FOM_S$ [f] | 116.4 | 0.57 | 162.8 | 80.8 | 19.2 | 10.3 | **337.84** | 140.27 | 7.2 | 2.3 | 0.16 |
| $FOM_L$ [g] | **71.5** | 0.03 | 18.1 | 9.6 | 7.6 | 3.9 | 112.61 | 28.05 | 21 | 1.1 | N/A |

[a] Excluding biasing transistors for fair comparison (biasing network is highly design-dependent and shared across OTAs). [b] SUB=subthreshold, SAT=saturation, GD=gate-driven, BD=bulk-driven. [c] not directly measurable, lower-bounded by 10-pF parasitic capacitance (pad, bonding, PCB), 10-fF from simulations. [d] avg over 10 samples, [e] percentage variability $\sigma/\mu$ over 10 samples. [f] $FOM_S = \frac{GBW \cdot C_{L,max}}{P_{DD}} \left[\frac{MHz \cdot pF}{\mu W}\right]$ [g] $FOM_L = \frac{SR \cdot C_{L,max}}{P_{DD}} \left[\frac{V/\mu s \cdot pF}{\mu W}\right]$

variability, which is 94.5-214X lower than prior 4-stage OTAs [5], [6], [9], [10]. Area is higher compared to other solutions, as expected from sub-threshold operation. At supply voltages lower than ~0.6 V and under rail-to-rail operation requirement, bulk-driven amplifiers are a valid approach at the expense of reduced gain [4], [8]. The PSRR measured at 1 kHz is 51 dB (58 dB from post-layout simulations). This value is mainly determined by the adoption of a more scaled technology and, hence, a reduced DC gain if compared with [5] (same topology).

From Table II, the OTA power efficiency under small- and large-signal condition is quantified by the well-known figures of merit $FOM_S$ and $FOM_L$. Compared to the state-of-the-art, the $FOM_S$ is 1.44-204X better than other 4-stage OTAs in spite of having the lowest load capacitance among 4-stage OTAs (which tends to make the FOM less favorable), except from [9] that is just 1.4X higher. However, when fairly compared to sub-1 V multi-stage OTAs, the proposed OTA improves the small-signal $FOM_S$ by 11.3-727X over prior art. Thanks to the adopted optimized design strategy, the large-signal figure of merit ($FOM_L$) is better than all prior 4-stage and sub-1V 2- and 3-stage OTA, representing an improvement by a factor of 3.4-65X, as expected from the considerations in Section III.

Note that although [7], [10] and [11] show a higher capacitive load range (30-200 pF minimum load), their gain-bandwidth product is reduced of a factor 6.4-96,000X. At the same time, this leads to a power consumption increase by 5.4X-214X, which makes these solutions unsuitable for low-power systems under wide load range conditions.

To summarize, the proposed 4-stage OTA entails minimal compensation complexity in terms of procedure and passives. The compensation network requires only a single Miller capacitor and an RC circuit, fundamentally addressing the well-known difficulty in compensating 4-stage OTAs, and bringing its complexity back to 3-stage. The OTA widens the load range to the 100X mark, thanks to the lower sensitivity of the phase margin to the load capacitance (easy calculations omitted). Its power efficiency when driving a 1-nF capacitive load outperforms state-of-the-art 4-stage OTAs by 4X in a large-signal condition, and by 11.3X in a small-signal condition for sub-1V OTAs. Accordingly, the proposed OTA and its compensation strategy are well suited for low-complexity, low-power and low-voltage targets, while efficiently driving loads and maintaining stability over an uncommonly wide range of load capacitances for wide adoption at low cost and power.